\nonstopmode

\documentclass[%
 aip,
rsi,%
 amsmath,amssymb,
 reprint,%
]{revtex4-1}

\usepackage{graphicx}
\usepackage{dcolumn}
\usepackage{bm}

\begin{document}
\title{Gigahertz optomechanical modulation by split-ring-resonator nanophotonic meta-atom arrays}
\author{Y. Imade}
\author{R. Ulbricht}
\altaffiliation{Current address: Division of Chemistry and Biological Chemistry, School of Physical and Mathematical Sciences, Nanyang Technological University, Singapore 637371,
Singapore}
\author{M. Tomoda}
\author{O. Matsuda}
\affiliation{Division of Applied Physics, Graduate School of Engineering, Hokkaido University, Sapporo 060-8628, Japan}

\author{G. Seniutinas}
\author{S. Juodkazis}
\affiliation{Centre for Micro-Photonics, Faculty of Engineering and Industrial Sciences, Swinburne University of Technology, Hawthorn, VIC 3122, Australia}

\author{O. B. Wright}
\email{olly@eng.hokudai.ac.jp}
\affiliation{Division of Applied Physics, Graduate School of Engineering, Hokkaido University, Sapporo 060-8628, Japan}

\date{\today}

\begin{abstract}
Using polarization-resolved transient reflection spectroscopy, we investigate the ultrafast modulation of light interacting with a metasurface consisting of coherently vibrating nanophotonic meta-atoms in the form of U-shaped split-ring resonators, that exhibit co-localized optical and mechanical resonances. With a two-dimensional square-lattice array of these resonators formed of gold on a glass substrate, we monitor the visible-pump-pulse induced gigahertz oscillations in intensity of reflected linearly-polarized infrared probe light pulses, modulated by the resonators effectively acting as miniature tuning forks. A multimodal vibrational response involving the opening and closing motion of the split rings is detected in this way. Numerical simulations of the associated transient deformations and strain fields elucidate the complex nanomechanical dynamics contributing to the ultrafast optical modulation, and point to the role of acousto-plasmonic interactions through the opening and closing motion of the SRR gaps as the dominant effect. Applications include ultrafast acoustooptic modulator design and sensing.
\end{abstract}

\maketitle

Electromagnetic metamaterials are artificial media composed of arrays of unit structures much smaller than the optical wavelength, and can exhibit non-intuitive properties such as negative refractive index\cite{shelby2001}, magnetism at optical frequencies\cite{linden2006photonic} and cloaking\cite{schurig2006metamaterial}. Such behavior is determined by the electrical or magnetic character of the constitutent meta-atoms. Following on from the first demonstration of negative permittivity and permeability at microwave frequencies\cite{Smith2000} in a hybrid split-ring resonator (SRR) and rod-array structure, a great deal of research on metamaterials  has been carried out up to optical frequencies, with potential applications in telecommunications and sensing techonologies, for example. Meta-atoms at optical frequencies have been proposed in many forms, such as three-dimensional (3D) fishnets\cite{valentine2008three}, cut-wire pair arrays\cite{dolling2005cut} as well as the generic split-ring resonators (SRR)\cite{enkrich2005magnetic}, to name only a few. At such frequencies plasmonic effects play an essential role in determining the effective parameters. 

Another promising avenue for applications involves the use of active metamaterials for wave control\cite{zheludev2012metamaterials}; at optical or infrared frequencies, for example, various methods of modulating the effective parameters of the metamaterial to this end have been demonstrated, such as photoswitching\cite{dani2009subpicosecond,xiao2010loss}, heating and cooling (i.e thermal control)\cite{liu2016thermochromic, ou2016giant, pang2017thermally}, the use of phase-change materials\cite{sautter2015active,yin2015active}, electrooptic materials\cite{yao2014electrically,valente2015magneto}, or coupling with microelectromechanical systems\cite{gutruf2015mechanically, valente2015magneto,ee2016tunable}. In particular, in the case of SRR-based metamaterials, active control of their optical or infrared properties has been conducted by many groups using photoswitching or electrooptic control\cite{lee2012switching, gu2012active, liu2015highly, meng2017hybrid}, phase-change materials\cite{samson2010metamaterial,kodama2016tunable}, or mechanical deformations\cite{pryce2010highly, ou2011reconfigurable, pitchappa2016active}, for example.

A different route to such modulation is the use of phonons, which promise ultrahigh frequency operation and control. Ulbricht et al. \cite{ulbricht2017} used GHz acoustic phonons to modulate the transmission of a metalayer consisting of an array of nanoholes in a gold film, and O'Brien et al.\cite{o2014ultrafast} made use of GHz acoustic phonons in nanoscale gold Swiss-cross arrays with different lengths of horizontal and vertical arms to modulate linearly polarized light. Shelton et al.\cite{shelton2011strong} demonstrated the coupling between gold SRR-array infrared resonances and THz optical phonons in thin dielectric layers, thereby modulating the infrared transmission spectrum, and Dong et al.\cite{dong2015} reported the modulation of light with GHz acoustic phonons in U-shaped nanowire arrays.  However, in spite of the intense interest in the fascinating properties of SRRs, to our knowledge GHz acoustic phonons have not been used to modulate the optical properties of SRR meta-atoms. In this paper we report on the GHz acoustic modulation of a metamaterial consisting of a 2D array of U-shaped nanoscale gold SRRs at near-infrared optical wavelengths using a combination of both experiment, based on a femtosecond polarization-resolved pump-probe technique, and numerical simulations of the transient deformation and strain fields.

\begin{figure}[ht]
  \centerline{\includegraphics [width= 7 cm ]{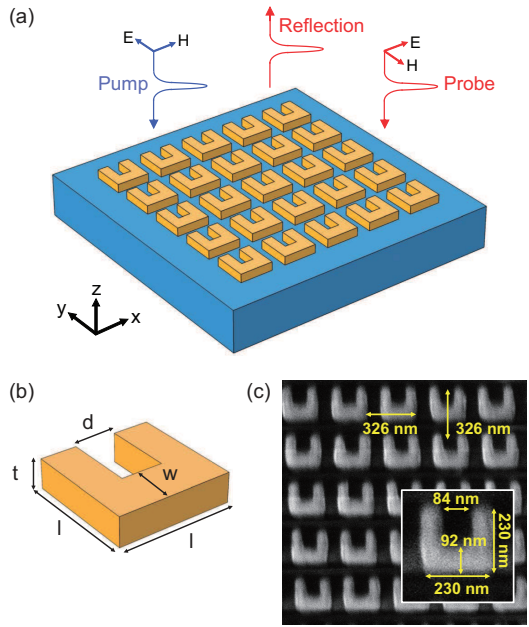}}
  \caption{(a) Schematic diagram of the sample configuration for horizontal probe polarization. (b) Definitions of the dimensions of the SRR.  (c) Scanning electron micrograph of the gold SRRs (of thickness $t = 60$ nm) fabricated on a BK7 glass substrate.}
  \label{Fig. 1}
\end{figure}

Our metamaterial consists of an array of sub-micron SRRs, as shown schematically in Figs.~1(a) and (b). The original Pendry-et-al.\cite{pendry1999magnetism} SRRs were proposed to achieve effective magnetic permeability through electromagnetic resonances at GHz electromagnetic frequencies. The fundamental magnetic resonance of an SRR can be approximated as an $LC$ circuit consisting of an inductance ($L$) and a capacitance ($C$). We have chosen U-shaped gold nanoscale SRRs of dimensions as shown in the inset of Fig.~1(c), so that their resonant frequency is in the near-infrared. Using the analytical $LC$ model of Linden et al.\cite{linden2006photonic}, we calculate this fundamental resonance to be at a wavelength of $\sim$1.5 $\mu$m. 
The SRR dimensions chosen, with reference to Fig.~1(b), are side $l$=230 nm, gap $d$=84 nm, bottom-width $w$=92 nm and thickness $t$=60 nm. They are arranged in a square lattice of pitch 326 nm, as shown by the electron micrograph in Fig.~1(b). The structures are patterned using electron-beam lithography and standard lift-off procedures\cite{Gervinskas2013}. A 0.5 mm thick slab of BK7 glass is used as a substrate, and a 2 nm Cr layer is incorporated to improve adhesion.

We first characterize the SRR array by normal-incidence white-light optical transmission spectra for horizontal and vertical polarizations, as shown in Fig.~2 for both experiment (solid red lines) and finite-element method (FEM) electromagnetic simulations (COMSOL Multiphysics, solid green lines) using periodic boundary conditions and literature values of the refractive indicies\cite{hagemann1975optical, Schott}. We also show for reference the simulated reflection spectrum (dotted blue lines) and absorption spectrum (dashed orange lines). Details of the simulations are given in the Supplementary Material. The horizontal optical polarization configuration of Fig.~2(a) is expected from analytical considerations to excite the fundamental magnetic resonance at $\sim$1.5 $\mu$m for our sample, but this resonance is out of the measured wavelength range. However, for this optical polarization we find first and second orders of plasmonic resonance at 808 nm and 572 nm in the simulations, labeled (1) and (2), respectively, in the reflection spectrum, and experimental transmission dips are correspondingly observed at wavelengths closely matching  those predicted in transmission. These resonances have been reported by other groups for similar U-shaped SRR arrays in the near infrared\cite{enkrich2005magnetic, rockstuhl2006resonances, linden2006photonic, corrigan2008optical}. Figure 2(b) shows the equivalent spectra for vertical polarization. This polarization does not excite a circulating current component in the split ring owing to symmetry. The directions and strengths of current flow at selected locations at a given time for the representative plasmonic resonances, calculated from electromagnetic simulations (COMSOL Multiphysics), are also shown in Fig.~2. (See the Supplementary Material for details and plots.) There is in general very good overall agreement between the experimental and simulated transmission spectra for both probe polarizations.

\begin{figure}[ht]
\centerline{\includegraphics [width= 6.5 cm ]{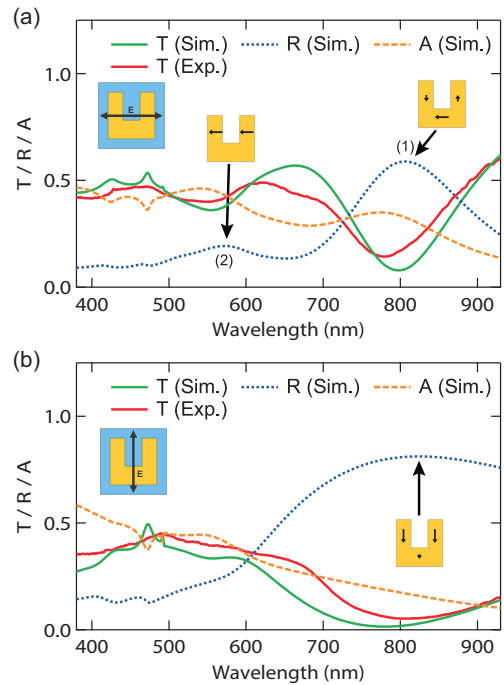}}
  \caption{(a), (b) show optical spectra for horizontal and vertical optical polarizations, respectively, for normal incidence intensity transmission (T): experiment (solid red lines) and simulation (solid green lines). Simulations of the intensity reflection (R)  spectrum (dotted blue lines) and absorption (A) spectrum (dashed orange lines) are also shown. Labels (1) and (2) in (a) refer to first- and second-order plasmonic resonances, with current flow as indicated in each case, calculated from electromagnetic FEM simulations (see Supplementary Material). A similar plasmonic resonance is indicated in (b).  Exp.: experiment. Sim.: simulation.}
  \label{Fig. 2}
\end{figure}

\begin{figure}[ht]
\centerline{\includegraphics [width= 7 cm ]{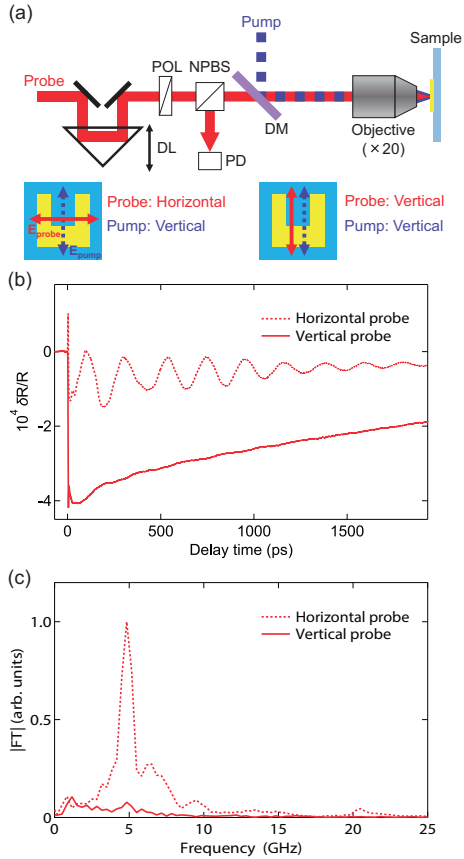}}
  \caption{(a) Schematic diagram of the experimental setup based on an optical pump-probe technique. DL: delay line, POL: polarizer, NPBS: non-polarizing beam spliter, PD: photodetector, and DM: dichroic mirror. Directions of linear polarization for the pump (dashed blue arrows) and probe (solid red arrows) beams are also shown.(b) Experimental pump-induced reflectivity changes of the sample vs delay time for horizontal (dotted red curves) and vertical (solid red curves) probe-beam polarizations, using a vertically-polarized pump beam, and (c) corresponding moduli of the temporal Fourier transforms ($|$FT$|$) plotted vs frequency on a normalized scale.}
  \label{Fig. 3}
\end{figure}

We use an optical pump-probe technique to generate GHz vibrational modes and to detect the modulated optical reflectance, as shown in Fig.~3(a). A mode-locked Ti:sapphire laser with a repetition rate of 80 MHz and optical pulse duration of $\sim$200 fs provides probe pulses at a wavelength of 800 nm as well as frequency-doubled pump pulses at a wavelength of 400 nm by use of a BBO (beta barium borate) crystal. At both 400 and 800 nm, gold is a good absorber with an optical absorption depth $\sim$15 nm\cite{hagemann1975optical}. The pump beam, chopped at 1 MHz by an acoustooptic modulator for lock-in detection, is focused onto the sample surface at normal incidence through a 20$\times$ objective lens to a spot diameter $\sim$6 $\mu$m FWHM (full width at half-maximum) with vertical linear polarization, as shown in Fig.~3(a) by the dashed (blue) arrow. Ultrafast electron diffusion in gold\cite{wright1994ultrafast} rapidly transfers energy from the optical absorption depth to the whole depth of the SRR. The resulting stress field coherently excites the vibrational modes of $\sim$300 SRRs, corresponding to near ${\bf{k}}$=0 acoustic wave vectors of the SRR array, which act as a phononic crystal. A single pump pulse has an energy of $\sim$50 pJ, which leads to acoustic strains up to $\sim$10$^{-4}$ or displacements of $\sim$20 pm, corresponding to transient temperature rises of $\sim$20 K. The probe pulses, of the same pulse energy, are passed through a motorized delay line, and are then focused at normal incidence to a similar spot size as the pump in order to monitor the transient reflectivity changes. The linear polarization of the probe beam is aligned alternatively horizontally and vertically, as shown in Fig.~3(a) (solid red arrows). The probe wavelength of 800 nm is conveniently set off the first plasmonic resonance of 780 nm for horizontal polarization (which produces electric-field localization in the SRR gap) to provide an enhanced sensitivity of the SRR reflectivity to deformation-induced changes in its geometry for this polarization. Variations in intensity $\sim$$10^{-4}$ of the probe beam reflected from the sample are monitored with a photodiode and a lock-in amplifer tuned to the chopping frequency.
\begin{figure}[ht]
\centerline{\includegraphics [width= 7 cm ]{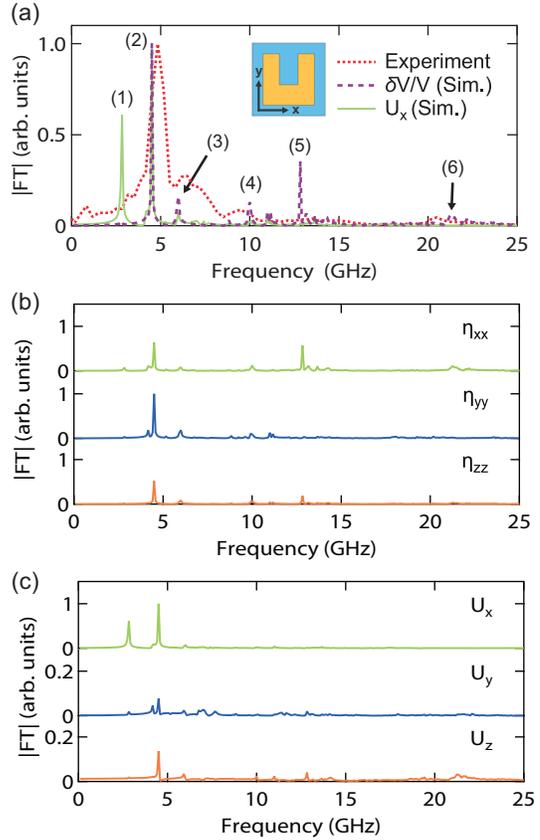}}
  \caption{(a) Normalized modulus of the temporal Fourier transform ($|$FT$|$) vs frequency for both experiment, for horizontal probe-beam polarization (dotted red line), and simulations for volumetric strain $\delta V/V$ (dashed purple line) and $x$-directed displacement $U_{x}$ (solid green line), both averaged over half of the top surface of the SRR. (b) $|$FT$|$ vs frequency for the simulated strain components: $\eta_{xx}$, $\eta_{yy}$ and $\eta_{zz}$, all averaged over half of the top surface of the SRR. (c) $|$FT$|$ vs frequency for the simulated displacement components: $U_{x}$, $U_{y}$ and $U_{z}$. Note the difference in vertical scales for $U_{y}$ and $U_{z}$ compared to $U_{x}$. Sim.: simulation.} 
  \label{Fig. 4}
\end{figure}

The measured relative reflectivity variations $\delta R(t)/R$ induced by the GHz vibrations in the SRR array are shown in Fig.~3(b) for horizontal (dotted line) and vertical (solid line) probe polarizations. At zero delay time, $\delta$$R/R$ shows rapid changes owing to electronic excitation and subsequent heating of the gold. After that, a complex oscillatory damped variation in $\delta R/R$ is evident, $\sim$10 times larger in the horizontal polarization case than in the vertical. We attribute this difference to the much sharper plasmon resonance close to the probe wavelength $\lambda$ for horizontal polarization compared to vertical (leading to a $\sim$10 times larger gradient $|dR/d\lambda|$\textemdash that governs the amplitude of the modulation in $\delta R/R$ when the plasmon resonance curve is shifted by strain or deformation\textemdash for horizontal polarization compared to vertical polarization at $\lambda$=800 nm).

The corresponding moduli of the temporal Fourier transforms ($|$FT$|$) of $\delta R/R$ (after subtracting the thermal background) are shown on a normalized scale vs frequency in Fig.~3(c). In particular a main resonance is evident in both polarizations at 4.8 GHz, and smaller resonances in the horizontal polarization at 3.2, 6.5. 9.5, 13.5 and 20.5 GHz appear. (The amplitude of the vertical probe polarization data is too small to extract other resonances.) Notably absent for both polarizations are Brillouin oscillations, that can arise from probe light scattered from coherent longitudinal strain pulses launched into the substrate. (These oscillations are expected to be very close in frequency to 22 GHz\cite{ulbricht2017}.) As we discuss in detail below, in the present experiment it turns out that plasmonic coupling dominates the detection process.

\begin{figure}[h!t]
\centerline{\includegraphics [width= 9 cm ]{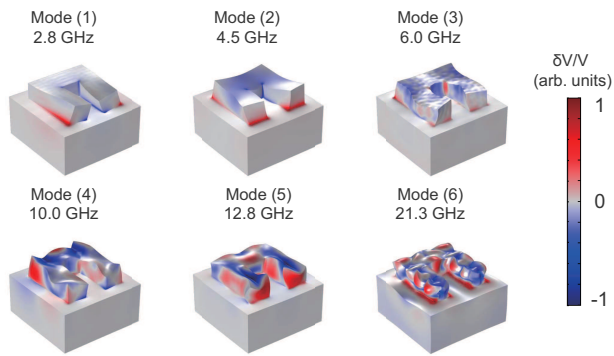}}
  \caption{Deformations and volumetric strain fields of the simulated vibrational modes at 2.8, 4.5, 6.0, 10.0, 12.8 and 21.3 GHz for a single unit cell of the structure. The amplitudes of deformations are greatly exaggerated compared to those in experiment. Animations can be viewed in the Supplementary Material.}
  \label{Fig. 5}
\end{figure}

In order to better understand these results, transient deformation and strain distributions in the SRR structure are calculated using frequency-domain FEM simulations (COMSOL, Multiphysics) with a mesh size of $\sim$8 nm, a time step of 1.0 ps, and a total calculation time of 12 ns. We implement periodic boundary conditions (BCs) with a unit cell consisting of a gold SRR (as defined by the experimental geometry) on a BK7 glass substrate section of dimensions 326 nm $\times$ 326 nm $\times$ 755 nm (ignoring the 2 nm Cr adhesion layer). Low-acoustic-reflection BCs are used over the bottom surface of the substrate. As an approximation to the thermoelastic excitation in gold, the whole SRR is subject to an initial isotropic stress that initiates an expansion. This assumption preserves the required left-right symmetry of the optical excitation in experiment with vertically-polarized normal optical incidence. In the simulations, literature values of longitudinal and shear sound velocities as well as densities of gold and BK7 are used: $v_{l} = 3240$ m/s, $v_{t} = 1200$ m/s, and $\rho$ = 19300 kg/m$^{3}$ and $v_{l} = 6050$ m/s, $v_{t} = 3680$ m/s, and $\rho$ = 2510 kg/m$^{3}$, respectively\cite{Schott, CRC}. (The accuracy of the
simulations were checked by using different discretizations.)

In order to obtain some quantitative parameters that can characterize the SRR vibrational motion, we plot vs frequency in Fig.~4(a) the normalized modulus of the temporal FT of the calculated volumetric strain $\delta V/V$ (dashed purple line) and displacement $U_{x}$ (solid green line), both averaged over half of the top surface of the SRR (i.e. exploiting the mirror symmetry), where $x$ and $y$ are the in-plane horizontal and vertical coordinates, respectively. For reference we also include the $|$FT$|$ of the experimental reflectivity change $\delta R/R$ (dotted red line) in the plot for horizontal probe polarization. Individual tensile strain component vibrational spectra $\eta_{xx}$, $\eta_{yy}$ and $\eta_{zz}$, averaged over half of the top surface of the SRR, are shown in Fig.~4(b), and all three displacement component spectra are shown in Fig.~4(c). In each case the relative values are accurately represented.

Several resonant frequencies are revealed that are very close to the experimental values. Since on resonance $U_{x}$ in the chosen frequency range is signficantly larger than $U_{y}$ and $U_{z}$ (see Fig.~4(c) noting the different vertical scales), we choose to reproduce its variation in Fig.~4(a) next to the experimental data. Deformations can contribute to the optical modulation in general in two ways: 1) by changing the probe plasmonic resonance frequency through, for example, a change in the SRR gap width; 2) by their role in changing the sample geometry, thus affecting the reflected angular distribution of the probe light, and leading to an intensity modulation owing to the finite optical solid angle defined by the collection optics. Effect 1) is likely to be more important in our SRR sample owing to the proximity of the horizontally-polarized probe wavelength to a plasmonic resonance. Likewise, strains can contribute to the optical modulation through strain-induced variations in the refractive index in the SRR or the substrate, although the photoelastic effect in gold at the 800 nm probe wavelength is known to be small, and has never been seen to give rise to any opto-acoustic interaction in experiments at such optical wavelengths\cite{ulbricht2017,garfinkel1966piezoreflectivity,he2017acoustic}. However, since similar strains are coupled to the glass substrate, the light reflected from the substrate around the SRR could well contribute to the optical modulation. For this reason we have included the above-mentioned volumetric strain $\delta V/V$=$\eta_{xx}$+$\eta_{yy}$+$\eta_{zz}$ in the plot of Fig.~4(a) for comparison. However, since $\eta_{xx}$ and $\eta_{yy}$ show similar amplitudes in Fig.~4(b) (and the same being true for the uncovered glass regions of the unit cell, as verified by simulation), one would expect the photoelastic effect in the glass substrate to modulate $\delta R/R$ with a similar amplitude for both horizontal and vertical probe polarizations. This can be seen from the following definitions:
\begin{equation}
\delta \epsilon_{xx}=P_{11} \eta_{xx} + P_{12} \eta_{yy} + P_{12} \eta_{zz},
\end{equation}
\begin{equation}
\delta \epsilon_{yy}=P_{12} \eta_{xx} + P_{11} \eta_{yy} + P_{12} \eta_{zz},
\end{equation}
where $\epsilon_{ij}$ are the dielectric constants and $P_{ij}$ the photoelastic constants ($P_{12}$/$P_{11}$$\sim$2 in glass\cite{dixon1967photoelastic}), showing that one expects the change $\delta \epsilon_{xx}$ (probed by horizontal polarization) to be of the same order as $\delta \epsilon_{yy}$ (probed by vertical polarization) when $\eta_{xx}$$\sim$$\eta_{yy}$, a situation that leads to a similar value of $\delta R/R$ for the two probe polarizations.
Since this is definitely not the case in experiment, one can conclude that it is likely to be the deformations of the SRR, i.e. particularly the component $U_{x}$, that dominate in the optical modulation. This conclusion is also consistent with the above-mentioned observed absence of Brillouin oscillations, which depend on the photoelastic effect.
\begin{figure}[h!t]
\centerline{\includegraphics [width= 7 cm ]{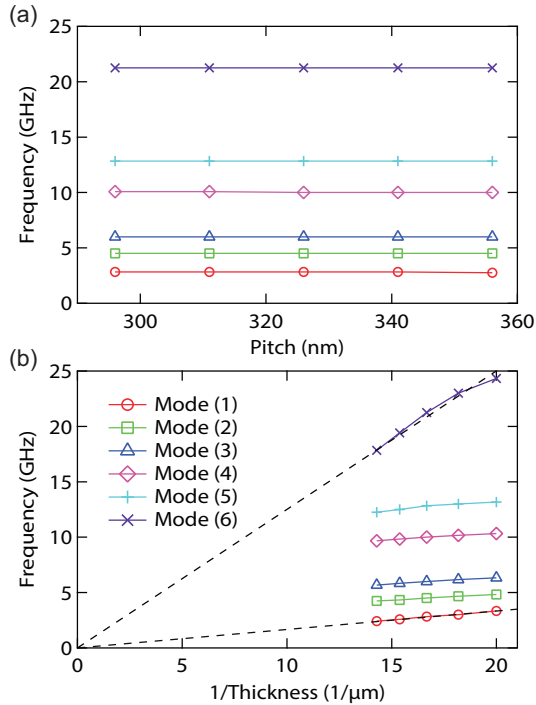}}
  \caption{(a) Plot of the frequencies of six simulated vibrational modes vs the SRR pitch $a$. (b) Plot vs inverse SRR thickness $1/t$. The dashed lines in (b) are fits in the form $f = K/t$, where $f$ is the mode frequency and $K$ is an adjustable constant.}
  \label{Fig. 6}
\end{figure}

The simulated FT spectra of both $\delta V/V$ and $U_{x}$ show easily recognizable resonances at 2.8, 4.5, 6.0, 10.0, 12.8 and 21.3 GHz. We have labeled as peaks (1)-(6) in Fig.~4(a) the ones that are close to maxima in experiment at 3.2, 4.8, 6.5. 9.5, 13.5 and 20.5 GHz, respectively, in the same figure. Whether a particular peak appears in experiment depends on the detailed thermoelastic coupling of the SRR to the substrate as well as on the specfic detection mechanism, so we do not expect the simulated peak heights to be the same as those in experiment; a comprehensive theory of the opto-acoustic nanoscale interaction is beyond the scope of this paper. The predicted peaks are much narrower than in experiment, presumably owing to variations in the SRR geometry (inhomogeneous broading) and to GHz ultrasonic attenuation in gold and in glass. The nature of these six labelled modes can be gleaned from the corresponding simulated deformations and strain distributions shown in Fig.~5. Resonance (1) corresponds to the fundamental tuning-fork-like mode, whereas modes (2)-(6) are of higher order. Their motion can be viewed as animations (see Supplementary Material), showing that all influence the SRR gap geometry. As prescribed by the reflection symmetry of the SRR in the bisector plane parallel to the $y$ axis, all the modes extracted show this required left-right mirror symmetry, and appear to all intents and purposes to be independent modes of isolated SRRs.

Concerning the large difference in experimental modulation amplitudes between horizontal and vertical probe polarizations, it is likely, as mentioned above, that plasmonic effects (i.e. perturbation of the enhanced $E$-fields) associated with the variation in the SRR gap, which are much more important for horizontal probe polarization, are mainly responsible. The tuning-fork-like vibration of the SRR ``prongs'' for the lower-order vibrational modes produces alternative blue-shifting (gap-opening) and red-shifting (gap-closing) of the plasmonic resonance, thus modulating the transient reflectivity\cite{huang2007,pryce2010highly,dong2015}.

To further investigate the origin of the vibrational modes dominant in the spectra for displacement and tensile strain, in the simulation we separately varied the SRR pitch $a$ the thickness $t$ around the chosen values, as shown in Fig.~6(a) and (b), respectively. The frequencies of mode (1)-(6) were independent of the pitch over the investigated range $a = 326 \pm 30$ nm, confirming our conclusion above that these modes can be treated to a good approximation as individual resonances of a single SRR in spite of being ${\bf{k}}$-near-zero collective modes of the phononic-crystal structure. In contrast to the variation with $a$, the frequencies of all six modes depend on thickness $t$ over the investigated range $t = 60 \pm 10$ nm, decreasing with increasing $t$ owing to the increased mass loading. Plots of the mode frequencies against $1/t$ in Fig.~6(b) show that for modes (1) and (6) the variation is approximately ${\propto}$ $1/t$, although we do not have a simple model to mimick this behaviour. Such variations of GHz mode frequency with nanoscale geometry have previously been investigated in phononic crystals (see, for example, Refs. \onlinecite{sakuma2012vibrational} and \onlinecite{robillard2008collective}).

In conclusion, we have measured the ultrafast acoustooptic response of SRR nanophotonic meta-atoms for the first time by means of femtosecond polarization-resolved transient reflection spectroscopy in the near-infrared. Ultrafast changes in optical reflectivity arise from GHz acoustic vibrations of a square lattice of suboptical-wavelength size thin gold U-shaped SRRs attached to a glass substrate. Individual vibrational modes of the SRRs in the frequency range $\sim$3-20 GHz are clearly identified to be photoexcited by the optical pump pulses and to give rise to a modulation in reflection $\delta R$ that is much larger for linear probe polarizations aligned across the SRR gap (i.e. horizontally rather than vertically). Simulations of the complex transient deformations and strains in the sample give reasonable overall agreement with the six resonant frequencies observed in experiment. Our analysis suggests that in this structure it is the deformations that tend to open and close the SRR gaps that contribute most to $\delta R$ through acousto-plasmonic effects rather than through the photoelastic effect, demonstrating the importance of the optomechanical interaction of the enhanced localized electric field in the gap with the GHz gap variations. In future, it would be interesting to determine how the acoustic modulation affects the angular distribution of optical intensity scattered from the SRR structures, both in reflection and transmission. Moreover, by tailoring the geometry and choice of material for the SRR, for example by extending the structure in 3D\cite{liu2008three}, it should be possible to enhance the acousto-plasmonic interaction, thereby opening the way for efficient ultrafast acoustic modulation using SRR meta-atoms. Our study also opens new vistas for the design of coherent phonon devices sensitive to variations in the phononic, electronic or thermal environment.

We are grateful to Kentaro Fujita for stimulating discussions. We acknowledge Grants-in-Aid for Scientific Research from the Ministry of Education, Culture, Sports, Science and Technology (MEXT) and well as support from the Japanese Society for the Promotion of Science (JSPS).

\clearpage

\setcounter{figure}{0}

\makeatletter 
\renewcommand{\thefigure}{S\@arabic\c@figure}
\makeatother

\begin{center}
\section*{Supplementary Material}
\end{center}

\section*{Optical 3D finite-element modelling}

The optical reflectance, transmittance and absorption spectra of the sample for normal incidence are simulated with a frequency-domain finite-element method (COMSOL Multiphysics) to solve the Maxwell equations. The unit cell consists of a BK7 glass substrate, a gold split-ring resonator (SRR) and the vacuum above them, as shown in Fig. S1. The SRR dimensions chosen, with reference to Fig. 1(a) of the main text, are side $l$=230 nm, gap $d$=84 nm, bottom-width $w$=92 nm and thickness $t$=60 nm. We implement periodic boundary conditions (BCs) with an unit cell of a gold SRR set on the interface between the vacuum and the BK7 glass substrate. S-parameter ports are used over two parallel planes, termed the input and output planes, defined to be 400 nm from this interface, and optical scattering BCs, set 100 nm outside these ports, are also used, as shown in Fig. S1.  A total of approximately 1,150,000 tetrahedral mesh elements are used for the discretization, of maximum side 63 nm in the vacuum, maximum
42 nm in the BK7 substrate and maximum 15 nm in the gold SRR. (These mesh sizes are six times smaller than the minimum simulated optical wavelength in each medium.) A finer mesh was used near the corners in the gold SRR, where the electric fields are particularly enhanced. The refractive indices of gold and BK7 glass are taken from the literature\cite{SCHOTT, hagemann1975optical}

 Simulations were performed independently with electric field {\bf{E}} polarized along the $x$ axis (horizontal polarization) and with {\bf{E}} polarized the $y$ axis (vertical polarization). 
The simulated reflectance, transmittance, and absorption for the two polarizations are plotted in Fig. 2(a) and (b) of the main text, respectively. Vector {\bf{E}}-field and current density {\bf{J}} plots at the optical wavelengths corresponding to the three peaks identified in Fig. 2(a) and (b) of the main text are shown in Fig. S2. 
\begin{figure}[h]
  \centerline{\includegraphics [width= 7 cm ]{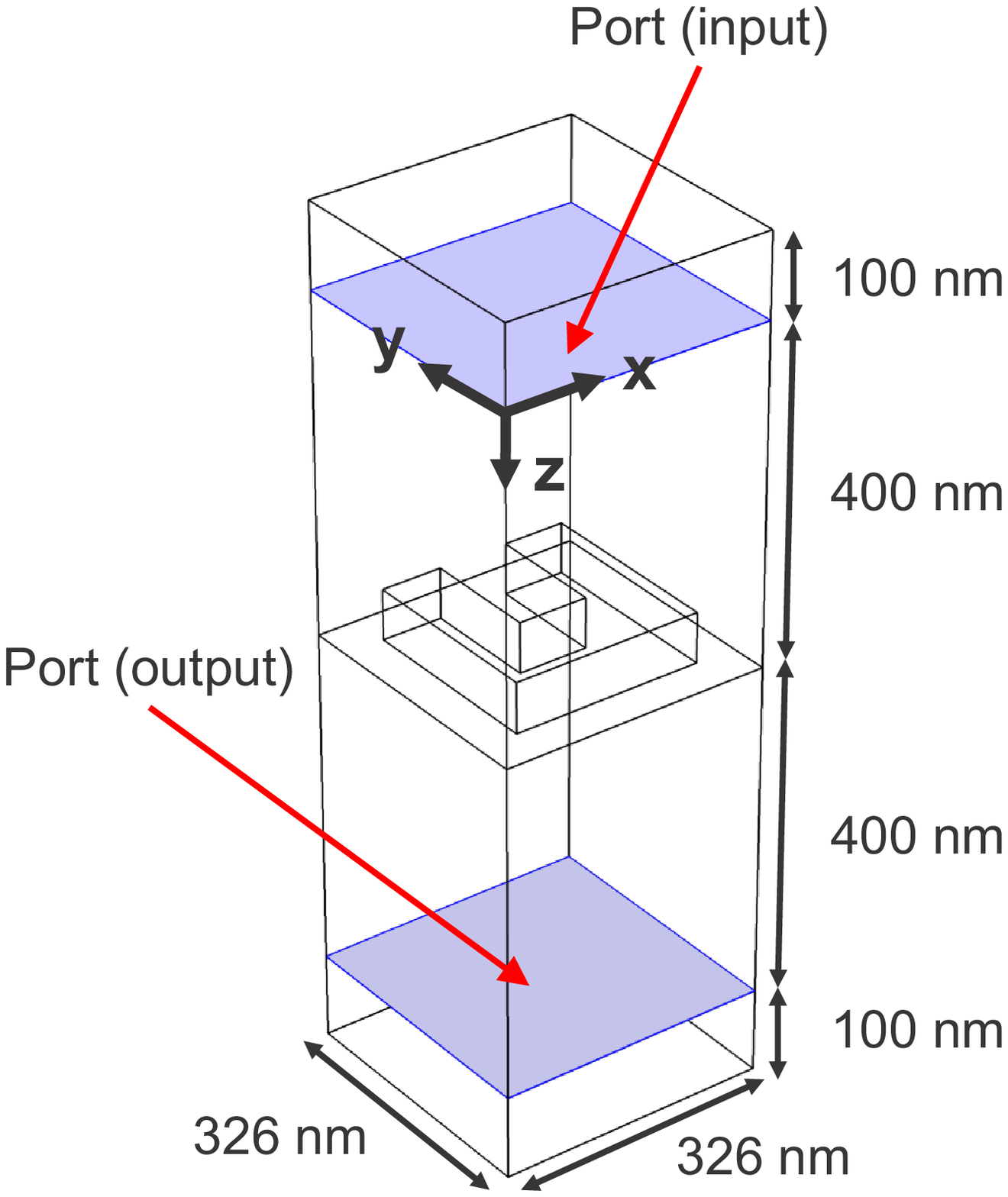}}
  \caption{Geometry of the SRR unit-cell structure used for electromagnetic simulations}
  \label{Fig. S1}
\end{figure} 
\begin{figure}[h]
  \centerline{\includegraphics [width= 7 cm ]{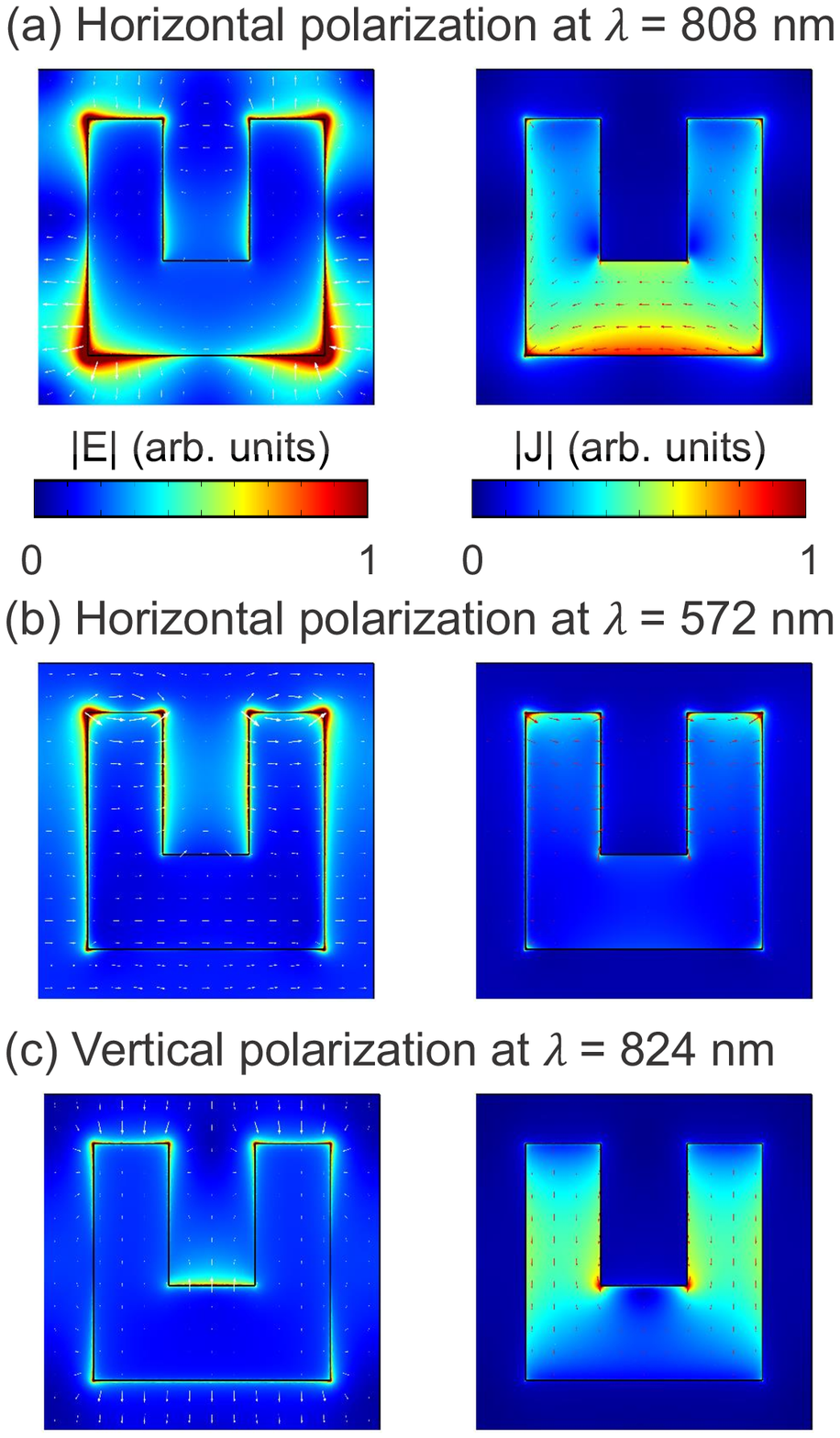}}
  \caption{Vector {\bf{E}}-field (left) and current density {\bf{J}} (right) plots for optical wavelengths corresponding to (a) peak (1) and (b) peak (2) in Fig. 2(a) of the main text, and (c) at the peak of wavelength 824 nm in Fig. 2(b) of the main text.}
  \label{Fig. S2}
\end{figure}

\end{document}